\documentclass[aps,twocolumn,superscriptaddress,showkeys,showpacs]{revtex4}
\usepackage[dvips]{graphicx}
\usepackage{amsfonts} 
\usepackage{amsmath,amssymb} 
\usepackage{latexsym} 
\usepackage{amssymb}
\usepackage{eurosym}
\usepackage{color}

\newcommand{\vt}{V}
\newcommand{\ports}{\pi}
\newcommand{\port}{\!:\!}
\newcommand{\dom}{\textrm{dom}}

\newcommand{\ov}{\overline}

\newcommand{\ket}[1]{| #1 \rangle}
\newcommand{\bra}[1]{\langle #1 |}
\newcommand{\braket}[2]{\langle #1 | #2 \rangle}
\newcommand{\trace}{\textrm{Tr}}

\newcommand{\joliH}{\mathcal{H}}

\newcommand{\pa}[1]{\left(#1\right)}

\newtheorem{Def}{Definition}
\newtheorem{Th}{Theorem}

\newtheorem{Cor}{Corollary}
\newtheorem{Pro}{Proposition}
\newtheorem{Rk}{Remark}

\newcommand{\couic}[1]{}

\begin{document}

\title{\begin{center}
Quantum Causal Graph Dynamics
\end{center}}

\begin{abstract}
Consider a graph having quantum systems lying at each node. Suppose that the whole thing evolves in discrete time steps, according to a global, unitary causal operator. By causal we mean that information can only propagate at a bounded speed, with respect to the distance given by the graph. Suppose, moreover, that the graph itself is subject to the evolution, and may be driven to be in a quantum superposition of graphs---in accordance to the superposition principle. We show that these unitary causal operators must decompose as a finite-depth circuit of local unitary gates. This unifies a result on Quantum Cellular Automata with another on Reversible Causal Graph Dynamics. Along the way we formalize a notion of causality which is valid in the context of quantum superpositions of time-varying graphs, and has a number of good properties. 
\end{abstract}



\author{Pablo Arrighi}
\email{pablo.arrighi@univ-amu.fr}
\affiliation{Aix-Marseille Univ., CNRS, LIF, Marseille and IXXI, Lyon, France}

\author{Simon Martiel}
\email{martiel@lsv.ens-cachan.fr}
\affiliation{INRIA Saclay, Team DEDUCTEAM, LSV Cachan, Cachan, France}

\maketitle

\section{Introduction}\label{sec:introduction}

\begin{center}
{\em Causality versus localizabity.}
\end{center}
 
Causality refers to the physical principle according to which information propagates at a bounded speed. Localizability refers to the principle that all must emerge constructively from underlying local mechanisms, that govern the interactions of close by systems.

Classically there may not be much difference between the two. Consider Cellular Automata (CA), for instance, i.e. a grid of cells, each of which may take one in a finite number of possible states. Causality in this context states that the next state of a cell must be a function of its current state and that of its neighbours. But then this update function readily provides us with the underlying local mechanism demanded by localizability. Things get much more involved if causality is relaxed to its topological characterization \cite{Hedlund}, and if the grid is relaxed to time-varying graphs \cite{ArrighiCGD,ArrighiRC}, a.k.a. for Causal Graph Dynamics (CGD).  Still, causality is shown to imply localizability.\\
In the reversible setting causality is no different, but localizability is more stringent, because the local mechanism must itself be reversible. Still it was shown that Reversible CA decompose as a finite-depth circuit of reversible, local gates \cite{KariBlock,KariCircuit,Durand-LoseBlock}. The same holds true for Reversible CGD \cite{ArrighiBRCGD} in spite of the dynamicity of the neighbourhood relation. In the probabilistic setting, however, the implication fails \cite{Henson,ArrighiPCA}.

It may therefore come as a surprise that, in the quantum setting, unitarity plus causality implies localizability. This was show successively for two systems \cite{Beckman}, three systems \cite{SchumacherWestmoreland}, a line of systems \cite{SchumacherWerner,ArrighiLATA,ArrighiIJUC} and eventually for an arbitrary fixed graph of systems \cite{ArrighiJCSS} --- encompassing Quantum CA.
In this paper we prove that, in the context of unitary evolutions of quantum superpositions of graphs, causality implies localizability. 

\begin{center}
{\em Quantum superpositons of graphs.}
\end{center}

Picture yourself a graph having quantum systems lying at each node. Suppose that the whole thing evolves in discrete time steps, according to a global unitary operator. But in such a way as to respect the graph: in one time step, information propagates from one node to another only if they are close by in the graph. This can all be defined and studied, these are unitary causal operators \cite{ArrighiJCSS}. But now, suppose that the graph itself is subject to the evolution, and gets driven to be in a quantum superposition of graphs---in accordance to the superposition principle. What does it mean to be causal in this strange context? When two nodes are now connected and disconnected, in a superposition, can they signal?

In this paper we propose and formalize a notion of causality in the context of quantum superpositions of time-varying graphs. The notion is well-behaved. In the quantum-but-fixed-topology regime, it specializes down to the more usual notion of causality used for Quantum Cellular Automata or in Algebraic Quantum Field Theory. In the classical-but-dynamical-topology regime, it specializes down to that used for Causal Graph Dynamics. It admits both a Shr\"odinger form, and a dual Heisenberg form, which remain equivalent. We do the same applies to the notion localizability. Even the basic operations of tensor product and partial trace demand slight generalizations in order to address this context. 

\begin{center}
{\em Motivations and result.}
\end{center}

Say that a Shr\"odinger cat is in a superposition of having fallen dead and being standing, alive. The cat's position changes the mass distribution in space, and so the curvature of space must also be in a superposition. What mathematical formalism can we use to describe this situation? Can we at least build a simple, discrete model that accounts for it? These sort of questioning have been at the heart of the research in Quantum Gravity. 

Quantum Graphity \cite{QuantumGraphity1,QuantumGraphity2}, for instance, considers a complete graph whose edges each carry a qubit that says whether the edge is active or not. The whole thing evolves in continuous-time, according to a Hamiltonian which is a sum of nearest-neighbours---in the sense of being connected by an active edge. Similarly Causal Dynamical Triangulations \cite{LollCDT} considers simplicial complexes evolving according to a path-integral formalism. So is the case of 
Loop Quantum Gravity \cite{RovelliLQG} in general, this time over a particular set of labelled graphs called spin foams. Much of the current research effort is dedicated towards understanding how an almost-flat space would emerge at large scale. We do not tackle this issue. 

\begin{figure}
\includegraphics[scale=0.7]{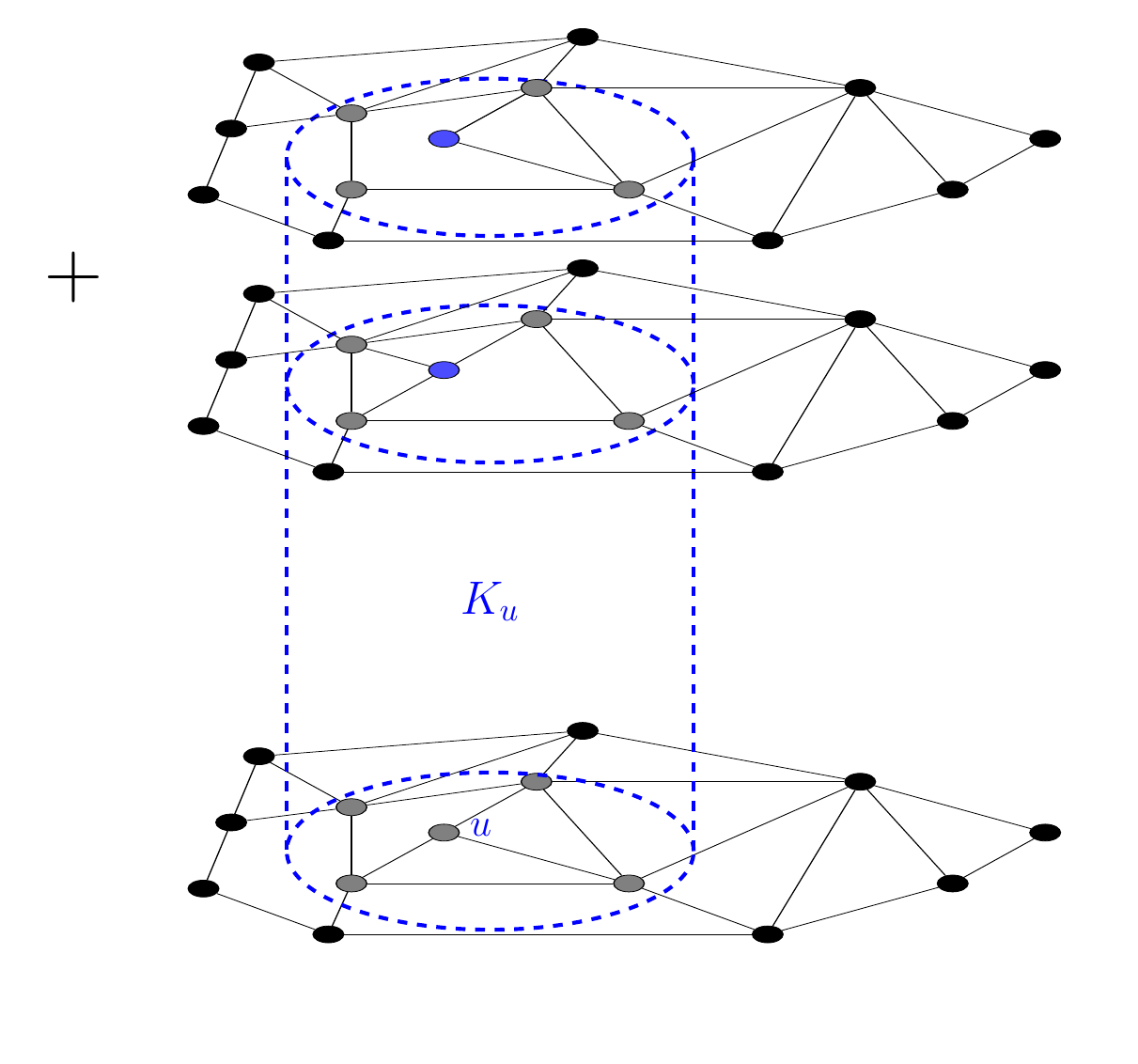}
\caption{\label{fig:K} $K_u$ is a local unitary gate: it acts locally upon vertex $u$ and its neighbours, and may create superpositions of graphs. Consider global, unitary causal operator, which takes quantum superpositions of entire graphs into quantum superpositions of entire graphs---but in such a way that information propagates from one node to another only if they are neighbours in the graph. In this paper we show that unitary causal operators always decompose as a finite-depth circuit of such local unitary gates.}
\end{figure}

What we provide is a discrete-time formalism for these quantum superpositions of time-varying graphs. Moreover, we provide a structure theorem which decomposes the global unitary operator $U$, into a finite-depth circuit of local unitary gates $\mu$ and $K$. This circuit can be described by a formula:	
$$U\ket{\psi} =(\prod  \mu_u)(\prod K_u)\ket{\psi},$$
whose meaning is illustrated in Fig. \ref{fig:K} and made clear in the text. This makes the model constructive and parametrizable. Moreover only the $K_u$'s depends on which $U$ we decide to implement that way, and these commute: $[K_u,K_v]=0$. Hence the first product, which is the relevant one, can be spectrally decomposed. 

\begin{center}
{\em Plan.}
\end{center}

In Sec. \ref{sec:defs} we formalize our state space: superpositions of graphs. In Sec. \ref{sec:tens} we adapt the notion of tensor and partial trace to this state space. In Sec. \ref{sec:caus} and \ref{sec:loc} we propose and formalize the notions causality and locality in this context, and prove several propositions of general interest. In Sec. \ref{sec:mainth} we prove our main result. Sec. \ref{sec:conclusion} provides a summary and perspectives.

\section{Mathematical preliminaries}
\label{sec:defs}

The graphs we consider are the usual, finite, undirected, bounded-degree graphs, but with three added twists:
\begin{itemize}
\item[$\bullet$] Edges are between ports of vertices, rather than vertices themselves, so that each vertex can distinguish its different neighbours, via the port that connects to it. 
\item[$\bullet$] The vertices are given labels taken in an alphabet, so that they may carry an internal state just like the cells of a Cellular Automaton. 
\item[$\bullet$] The labelling function is partial, so that we may express our partial knowledge about part of a graph.
\end{itemize} 
\begin{Def}[Graph]\label{def:graphs}
Consider $V$ a countable set of vertices and $\pi$ a finite set of ports.
A {\em graph} $G$ is given by 
\begin{itemize}
\item[$\bullet$] A finite subset $\vt(G)$ of $V$, whose elements are called {\em vertices}.
\item[$\bullet$] A set $E(G)$ of non-intersecting two-elements subsets of $\vt(G)\times\ports$, whose elements are called {\em edges}. In other words an edge $e$ is of the form $\{u \port a, v \port b\}$, and $\forall e,e'\in E(G), e\cap e'\neq \varnothing \Rightarrow e=e'$. 
\end{itemize}
The {\em set of all graphs} with vertices in $V$ and ports in $\ports$ is written ${\cal G}_{V,\ports}$. To ease notations, we write $v \in G$ for $v \in \vt(G)$. We write $\varnothing$ for the empty graph.
\end{Def}
\begin{Def}[Labelled graph]
A {\em labelled graph} is a triple $(G,\sigma)$, also denoted simply $G$ when unambiguous, where $G$ is a graph, and $\sigma$ is a partial function from $V(G)$ to a finite, or countable set $\Sigma$. The {\em set of all labelled graphs} with vertices in $V$, ports in $\ports$, and states in $\Sigma$ is written ${\cal G}_{V,\Sigma,\ports}$, or simply ${\cal G}$. 
\end{Def}
We need a notion of {\em union} of graphs, and for this purpose we need a notion of {\em consistency} between the operands of the union, so as at to make sure that both graphs ``do not disagree''. 
\begin{Def} [Consistent]
Two graphs $G$ and $H$ in ${\cal G}$ are {\em consistent} if and only if:
\begin{itemize}
\item over the set $W=(\vt(G)\cap \vt(H))$ the partial functions $\sigma_{G}$ and $\sigma_{H}$ agree when they are both defined, meaning that: $\forall u\in W$,
$$\big[u\in\dom(\sigma_G)\wedge u\in\dom(\sigma_H)\Rightarrow \sigma_G(u)=\sigma_H(u)\big]$$
where $\dom(\sigma)$ stands for the domain of $\sigma$.
\item the edges adjacent to vertices of $W$ in $G$ and $H$ agree, meaning that: $\forall u\port i\in (W\port\ports)$, $\forall v\port j\in V(G)\port\ports$, $\forall v'\port j' \in V(H)\port\ports$:
\begin{align*}
\big[\{u\port i, v\port j\}\in E(G) \wedge \{u\port i,& v'\port j'\}\in E(H)\\ &\Rightarrow (v=v'\wedge j=j')\big]
\end{align*}
\end{itemize}
\end{Def}

\begin{Def} [Union]
Consider two consistent graphs $G$ and $H$, we define the graph
$G\cup H$ to be the graph:
\begin{itemize}
\item whose set of vertices $\vt(G\cup H)$ is $\vt(G)\cup \vt(H)$. 
\item whose partial function $\sigma_{G\cup H}$ has domain $\dom(\sigma_{G})\cup\dom(\sigma_{H})$ and coincides with $\sigma_{G}$ (resp. $\sigma_H$) over $\dom(\sigma_G)$ (resp. $\dom(\sigma_H)$).
\end{itemize}
\end{Def}

We will also need ways of taking subgraphs that induced by the neighbours of a vertex. 
\begin{Def}[Disks]\label{def:disks}
Consider $G\in {\cal G}$, and $S\subset V$. We write $D(G)^r_S$ for the {\em radius $r$ neighbours of $S$ in $G$}, i.e. all those vertices which can be reached in $r$ steps, or less, following edges of $G$, starting from a vertex in $S$. This includes $S$. The {\em border vertices of $S$}, on the other hand, are the radius $1$ neighbours of $S$ in $G$ which do not lie in $S$.\\
Now consider $D\subset V$, and $\ov{D}=V\setminus D$. We write $G_D$ for the subgraph induced by $D$ and its border vertices, all labellings included except those of the border vertices. We write $\ov{G}_D$ for the subgraph induced by $\ov{D}$, all labellings included.\\
In the special case where $D$ is $D(G)^r_v$ the set of radius $r$ neighbours of vertex $v$ in $G$, we simply write $G^r_v$ instead of $G_{D(G)^r_v}$, and refer to this as a {\em disk}. Similarly we write $\ov{G}^r_v$ instead of $\ov{G}_{D(G)^r_v}$ and refer to this as the {\em complement of a disk}.
\end{Def}
\begin{Rk} Notice that for all $G$ and $D\subset V$, $G=G_D\cup\ov{G}_D$. Indeed, the decomposition does not miss any out edge between $D$ and $\ov{D}$, as these belong to $G_D$.
\end{Rk}

Having defined the set of labelled graphs, which is infinite but countable, we can readily use it as the canonical basis for a Hilbert space of quantum superpositions of graphs.
\begin{Def}[Superpositions of configurations]~\label{def:HCfbis}\\ 
We define ${\cal H}_{\cal G}$ be the Hilbert space of labelled graphs, as follows: to each labelled graph $G$ is associated a unit vector $\ket{G}$, such that the family $\pa{\ket{G}}_{G\in{\cal G}}$ is the canonical orthonormal basis of ${\cal H}_{\cal G}$. A \emph{state vector} is a unit vector $\ket{\psi}$ in ${\cal H}_{\cal G}$. A \emph{state} is a trace-one positive operator $\rho$ over ${\cal H}_{\cal G}$. To ease notations, we write $\joliH$ instead of ${\cal H}_{\cal G}$.
\end{Def}
\couic{
\begin{Def}[Isomorphism]
An isomorphism is specified by a bijection $R$ from $V$ to $V$ and acts on a graph $G$ of ${\cal G}$ as follow:
\begin{itemize}
\item $V(R(G))=R(V(G))$
\item $\{u:k,v:l\}\in E(G) \Leftrightarrow \{R(u):k,R(v):l\}\in E(R(G))$
\end{itemize}
The definition naturally extends to labelled graphs. It is linearly extended to any state of $\joliH$, thus defining a unitary operator.
\end{Def}

\begin{Def}[Dynamics]~\label{def:dyn}\\ 
A linear operator $U:\joliH\longrightarrow\joliH$ is said to be a \emph{dynamics} if and only if for any isomorphism $R$, we have $[U,R]=0$.
\end{Def}
}

\begin{Def}[Vertex preserving]~\label{def:vp}\\ 
A linear operator $U:\joliH\longrightarrow\joliH$ is said to be vertex preserving if and only if for any graph $G\in {\cal G}$, we have that if $U\ket{G}=\sum_i \alpha_i\ket{G(i)}$, then for all $i$, $V(G(i)) = V(G)$.
\end{Def}




\section{Generalized tensors and traces}\label{sec:tens}

In quantum theory, the tensor product is the basic operation used to mathematically represent the joint system of two systems next to one another. Here the systems are lie at vertices of a graph, and so we need to say how they connect to each other. Moreover, the graph itself may be in a superposition, it forms part of the state space. We need to adapt tensor products to this context.
\begin{Def}[Generalized Tensors]
We write $\ket{G}\otimes\ket{H}=\ket{G}\ket{H}$ for $\ket{G\cup H}$. The tensor product definition is then bilinearly extended to pairwise consistent superpositions of state vectors in ${\cal H}$.
\end{Def}
{\em Comment}. Usually a tensor product takes in states $\ket{\psi}^A$ and $\ket{\phi}^B$ from two non-overlapping systems $A$ and $B$, to produce $\ket{\psi}^A\ket{\phi}^B$. However, consider a common subsystem $C$ and input states from overlapping systems $AC$ and $BC$. If we demand that they have that the particular form $\ket{\psi}^A\ket{\vartheta}^C$ and $\ket{\phi}^B\ket{\vartheta}^C$, then we can naturally extend the tensor product to produce $\ket{\psi}^A\ket{\phi}^B\ket{\vartheta}^C$. This is what the above definition does: the consistency requirement amounts to imposing agreement upon the common subsystem.

\noindent Now, if $\rho$ captures the state of an entire system, then $\rho^r_v$ stands for the state of the neighbours of $v$:
\begin{Def}[Generalized partial trace]~\label{def:Tr}\\ 
Consider $\rho=\ket{G}\bra{H}$ over $\joliH$, with $G,H \in {\cal G}$. 
Let $D=D(G)^r_v\cup D(H)^r_v$. We define its partial trace $\rho^r_v$ to be 
$$\ket{G_D}\bra{H_D}\braket{\ov{H}_D}{\ov{G}_D}$$
The definition is linearly extended to any state over $\joliH$. 
\end{Def}
\begin{Rk} From the previous remark, we get that $\trace(\rho)=\trace(\rho^r_v)$.
\end{Rk}
{\em Comment}. On the one hand, the above definition is a straightforward extension of the usual $\trace_B(\ket{ij}^{AB}\bra{kl})=\ket{i}^A\bra{k}\braket{j}{l}$ formula. On the other hand, the definition intends to let the system $A$ be the disk of radius $r$ centered on $v$---but this a priori is an unclear notion for quantum superpositions of graphs. This issue is solved by addressing the basic case, first, and then extending to superpositions.

\section{Causality}\label{sec:caus}

A fundamental symmetry of physics is causality, meaning that information propagates at a bounded speed. In discrete space and time, this means that in order to know the next state and connectivity of a vertex $v$, we only need to know that of its neighbours at the previous time step:

\begin{Def}[Causality]~\label{def:causality}\\ 
A linear operator $U:\joliH\longrightarrow\joliH$ is said to be \emph{causal} if and only if for all $m\geq 0$ there exists $n\geq 0$ such that for any state $\rho$ over $\joliH$, and for any $v\in V$, we have
\begin{align}
(U\rho U^\dagger)^m_v=(U\rho^n_v U^\dagger)^m_v. \label{eq:causality}
\end{align}
\end{Def}

{\em Comment}. The above is a direct translation of causality, as expressed with our generalized partial trace. Notice how, in the case of basic states, this definition of causality specializes into the usual notion of causality as in classical Causal Graph Dynamics \cite{ArrighiCayley,ArrighiRC}. Notice also how, in the case of fixed graphs whose nodes are labelled by quantum states, this definition of causality again specializes into that used for Quantum Cellular Automata \cite{ArrighiUCAUSAL,ArrighiJCSS}. What happens in the grey zone of quantum superpositions of graphs may seem a little wilder, but in the end it is just the linear extension of these notions. 

The following proposition will turn out useful.
\begin{Pro}[Tensorial extension]\label{pro:tensext}
Given a causal unitary operator $U$ over ${\cal H}$, 
\begin{align*}
U'':{\cal H}\otimes{\cal H}&\longrightarrow{\cal H}\otimes{\cal H}\\
U''&=U\otimes I
\end{align*}
is causal. Here, ${\cal H} \otimes{\cal H} \equiv {\cal H}_{{\cal G}^2} \equiv {\cal H}_{{\cal G} \uplus {\cal G}}  $.
\end{Pro}

\noindent \textbf{Proof.} 
\begin{align*}
(U''\rho' U''^\dagger)^m_v=&(U''(\sum_i\rho(i)\otimes\tau(i)) U''^\dagger)^m_v\\
=&\sum_i (U\rho(i)U^\dagger)^m_v\otimes\tau(i)^m_v\\
=&\sum_i (U\rho(i)^n_v U^\dagger)^m_v\otimes(\tau(i)^n_v)^m_v\\
=&(U''\sum_i(\rho(i)\otimes\tau(i))^n_v) U''^\dagger)^m_v\\
=&(U''\rho'^n_v U''^\dagger)^m_v
\end{align*}
~\hfill$\Box$

\section{Locality}\label{sec:loc}

Causal operators change the entire graph in one go. The word causal there
refers to the fact that information does not propagate too fast. Local operations,
on the other hand, act just in one bounded region of the graph, leaving the rest
unchanged:
\begin{Def}[Localization]\label{localizationdef}
A linear operator $A:\joliH\longrightarrow\joliH$ is said to be $r$-\emph{localized} upon $v\in V$ if and only if for any $G,H \in {\cal G}$, we have that 
\begin{align}
\bra{H}A\ket{G}&=\bra{H_D}A\ket{G_D}\braket{\ov{H}_D}{\ov{G}_D} \label{eq:locality}
 \end{align}
with $D=D(G)^r_v\cup D(H)^r_v$. 
\end{Def}
In other words, $A$ only changes $v$ and its neighbours, and only requires knowledge of $v$ and its neighbours to do that. Thought of as a measurement, $A$ is only sensitive to changes in $v$ and its neighbours:
\begin{Pro}[Dual localization] \label{prop:dualloc}
Let $A$ be an $r$-local linear operator, upon a vertex $v$. This is equivalent to saying that for any two states $\rho, \rho'$ over $\joliH$, we have that $\rho^r_v=(\rho')^r_v$ entails that $\trace(A\rho)=\trace(A\rho')$. 
\end{Pro}
$[\Rightarrow]$. Suppose that $A$ is $r$-localized upon a vertex $v\in V$. For any $\rho=\ket{G}\bra{H}$ let $D=D(G)^r_v\cup D(H)^r_v$. We have:
\begin{align*}
\trace\pa{A \rho}&=\bra{H}A\ket{G}\\
&= \bra{H_D}A\ket{G_D}\braket{\ov{H}_D}{\ov{G}_D}\\
&= \trace\pa{A\ket{G_D}\bra{H_D}\braket{\ov{H}_D}{\ov{G}_D}}\\
&= \trace\pa{A\rho^r_v}
\end{align*}
Assuming $\rho^r_v={\rho'}^r_v$ yields $\trace\pa{A \rho}=\trace\pa{A\rho^r_v}=\trace\pa{A\rho'^r_v}=\trace\pa{A \rho'}$.\\

$[\Leftarrow]$. Let us assume that $A$ is dually $r$-local around $v$.  For any $\rho=\ket{G}\bra{H}$ Let $D=D(G)^r_v\cup D(H)^r_v$. We have:
\begin{align*}
 \bra{H}A\ket{G}&=\trace\pa{A\rho}\\
 &=\trace\pa{A\rho^r_v}\;\textrm{using dual localization.}\\
 &=\trace\pa{A\ket{G_D}\bra{H_D}\braket{\ov{H}_D}{\ov{G}_D}}\\
 &=\bra{H_D}A\ket{G_D}\braket{\ov{H}_D}{\ov{G}_D}
\end{align*}

Now that we have a notion of locality, we can rephrase that of causality in the Heisenberg picture, which is the more traditional one in algebraic quantum field theory for instance. 
\begin{Pro}[Dual causality]~\\ \label{prop:dual}
Let $U$ be a causal linear operator. This is equivalent to saying that for every operator $A$ $m$-localized upon vertex $v$, then $U^\dagger AU$ is $n$-localized upon vertex $v$.
\end{Pro}
\textbf{Proof.} 
$[\Rightarrow]$. Suppose causality and let $A$ be an operator $m$-localized upon vertex $v$. Let $n$ be that from Def. \ref{def:causality}. For every pair of states $\rho$ and $\rho'$ such that 
$\rho^n_v={\rho'}^n_v$, we have $\pa{U\rho U^\dagger}^m_v=\pa{U\rho' U^\dagger}^m_v$ and hence $\trace\pa{AU\rho U^\dagger}=\trace\pa{AU\rho' U^\dagger}$. We thus 
get $\trace\pa{U^\dagger AU\rho}=\trace\pa{U^\dagger AU\rho'}$. Since 
this equality holds for every $\rho$ and $\rho'$ such that 
$\rho^n_v={\rho'}^n_v$, we have that $U^\dagger AU$ is $n$-localized.\\
$[\Leftarrow]$. Suppose dual causality and $\rho^n_v={\rho'}^n_v$. 
Then, for every operator $B$ $n$-localized upon $v$, 
$\trace\pa{B\rho}=\trace\pa{B\rho'}$, and so for every operator $A$ $m$-localized upon vertex $v$, we get: 
$\trace\pa{AU\rho U^\dagger}=\trace\pa{U^\dagger 
AU\rho}=\trace\pa{U^\dagger AU\rho'}
=\trace\pa{AU\rho'U^\dagger}.$
This entails $\pa{U\rho U^\dagger}^m_v=\pa{U\rho' U^\dagger}^m_v$.\hfill $\Box$\\

\section{Representation theorem}\label{sec:mainth}
The goal of this section is to achieve a representation of causal operators as a bounded-depth circuit of local unitary operators. The idea of this construction is to proceed by local updates. We will construct local operators $K_u$ updating only the neighbourhood of a vertex $u$. All these $K_u$ will be local unitary operators and will commute with each other. To do so, we generalize the construction presented in \cite{ArrighiBRCGD}: the local update $K_u$ consists in applying the causal operator $U$,  ``putting aside'' vertex $u$ from the graph, and applying the inverse operator $U^\dagger$.

First, we construct an appropriate state space, allowing us to `mark' vertices in order to ``put them aside'' as computed. 

\begin{Def}[Marked graphs and space]
Given a set of graphs ${\cal G}={\cal G}_{V,\Sigma,\pi}$, consider the set of graphs ${\cal G}_{V,\Sigma',\pi'}$ with $\Sigma'=\Sigma\times\{0,1\}$ and $\pi'=\pi\times\{0,1\}$. We define the set of marked graphs ${\cal G}'$, to be the subset of ${\cal G}_{V,\Sigma',\pi'}$ such that for all graph $G\in {\cal G}'$, for all vertex $u\in G$ with $\sigma_G(u)=(x,a)$ and $\{u:(i,b),v:(j,c)\}\in G$, we have $a=c$. We denote by ${\cal H}'$ the state space whose basis vectors are the marked graphs ${\cal G}'$. Given a graph $G\in \mathcal{G}$, it is naturally identified with the same graph in ${\cal G}'$ with all marks set to $0$.
\end{Def}

The following definition introduces our marking mechanism.

\begin{Def}[Mark operator]
Given a set of labels $\Sigma'=\Sigma\times\{0,1\}$ and a set of ports $\pi'=\pi\times\{0,1\}$, we define the marking operation $\mu(.)$ over labels and ports as toggling the bit in the second component:
\begin{itemize}
\item $\forall (x,a)\in \Sigma', \mu(x,a)=(x,1-a)$
\item $\forall (i,a)\in \pi', \mu(i,a)=(i,1-a)$
\end{itemize}
Then, we define the mark operation $\mu_u$ over marked graphs, as attempting to mark the label of vertex $u$ and its opposite ports, if this will not create conflicts between ports. More formally, given a graph $G$ in ${\cal G}'$, we define the mark operation, $\mu_u:{\cal G}'\rightarrow {\cal G}'$ as follows:
\begin{itemize}
\item if $\exists v,w\in G ,i,j\in \pi'$ such that $\{u \port i ,  v \port j\}\in G$ and $\{ v\port \mu(j) ,w \port k\}\in G$ then $\mu_u G=G$
\item else
\begin{itemize}
\item[$\bullet$] $\sigma_{\mu_u G}(u)=\mu(\sigma_{G}(u))$
\item[$\bullet$] For all $i,j\in\pi'$,  $\{u\port \mu(i),u\port \mu(j)\} \in \mu_u G$ if and only if $\{u\port i,u\port j\} \in G$.
\item[$\bullet$] For all $v\in G$ with $v\neq u$ and $i,j\in\pi'$,  $\{u\port i,v\port \mu(j)\} \in \mu_u G$ if and only if $\{u\port i,v\port j\} \in G$.
\end{itemize}
with the rest of the graph $G$ left unchanged.
\end{itemize}
Finally,
$\mu_u$ is linearly extended to become a unitary operator over ${\cal H}'$. Moreover each $\mu_u$ is $1$-localized and commutes with $\mu_v$ for all $v$.
\end{Def}

\noindent\textbf{Soundness.} As $\mu_u$ specifies a bijection over the set of graphs ${\cal G}'$, its linear extension to ${\cal H}'$ is unitary. Moreover, $\mu_u$ only changes the label of vertex $u$ and the ports of its adjacent edges, but it does so conditionally upon the edges of the neighbours, which makes it $1$-localized. Finally, when $u\neq v$, then $\mu_u$ and $\mu_v$ either both act independently upon disjoint labels and ports, or they are both the identity---hence they commute.


It turns out that any causal operator admits an extension that is compatible with these marks.

\begin{Pro}[Marked extension]
Given a vertex-preserving causal unitary operator $U$ over ${\cal H}$, there exists a vertex-preserving causal unitary operator $U'$ over ${\cal H}'$ such that:
\begin{align*}
\forall G\in {\cal G}_{V,\Sigma\times\{0\},\pi\times\{0\}},&\quad U'\ket{G}=U\ket{G}\\
\forall G\in {\cal G}_{V,\Sigma\times\{1\},\pi\times\{1\}},&\quad U'\ket{G}=\ket{G}
\end{align*}
\end{Pro}
\noindent \textbf{Proof.} Consider
\begin{align*}
U'':{\cal H}\otimes{\cal H} &\longrightarrow {\cal H}\otimes{\cal H}\\
U''&=U\otimes I
\end{align*}
instead---which is causal by Proposition \ref{pro:tensext}. Notice that ${\cal H}\otimes{\cal H}\equiv {\cal H}_{{\cal G}^2}$. We now consider
\begin{align*}
\varphi:{\cal G}'&\longrightarrow{\cal G}^2\\
G&\mapsto (\ov{G}_M,\mu_M G_M)
\end{align*}
with $M$ the marked vertices of $G$. The function $\varphi$ is injective since if $\varphi (G')=(G,H)$ then we can recover $G'$ as $G\cup (\mu_M H)$ with $M=V(H)\backslash V(G)$. Let $S=\varphi(\cal{G}')$ be the image of $\varphi$.

Notice that, if $(X,H) \in S$ and $Y\in {\cal G}$ is such that $V(X)=V(Y)$, then $(Y,H)\in S$. Indeed, if $(X,H)$ has antecedent $X'=X \cup (\mu_M H)$ then the union $Y'=Y \cup (\mu_M H)$ is well-defined as $\mu_M H$ does not specify any internal state or connectivity in $\pi\times\{0\}$ over $V(H) \cap V(X)=V(H) \cap V(Y)$. Moreover, $\varphi(Y')=(Y,H)$ because since $V(X)=V(Y)$, $\ov{Y'}_M = Y$ and $\mu_M Y'_M=\mu_M X'_M=H$.

The subspace ${\cal H}_S$ is stable under $U''$. Indeed, consider $(G,H)\in S$
\begin{align*}
U''(\ket{G}\otimes \ket{H}) &= (U \ket{G})\otimes \ket{H} \\
&= \sum \alpha_i\ket{G(i)}\otimes\ket{H} 
\end{align*}
and notice that for all $i$, $(G(i),H)\in S$ because $V(G(i))= V(G)$.

Next, take $U'$ to be the restriction of $U''$ to $S$:
\begin{align*}
U':{\cal H}'&\longrightarrow {\cal H}'\\
\ket{G'}&\mapsto (\varphi^{\dagger} \circ U''\circ \varphi) \ket{G'}
\end{align*}
where $\varphi$ is linearly extended to be a unitary operator from ${\cal H}'$ to ${\cal H}_S$.
We have the two requested properties. Indeed, $\forall G\in {\cal G}_{V,\Sigma\times\{0\},\pi\times\{0\}}$,
\begin{align*}
U'\ket{G}&=\varphi^\dagger(U\otimes I)(\ket{G}\otimes\ket{\varnothing})\\
&=\varphi^\dagger(U\ket{G}\otimes\ket{\varnothing})\\
&=U\ket{G}
\end{align*}
and $\forall G\in {\cal G}_{V,\Sigma\times\{1\},\pi\times\{1\}}$,
\begin{align*}
U'\ket{G}&=\varphi^\dagger(U\otimes I)(\ket{\varnothing}\otimes\ket{\mu_{V(G)}\,G})\\
&=\varphi^\dagger(\ket{\varnothing}\otimes\ket{\mu_{V(G)}\,G})\\
&=\ket{G}
\end{align*}
Causality and vertex preservation are inherited from $U''$.
~\hfill$\Box$

The following theorem is our main contribution:
\begin{Th}[Structure theorem]~\\ \label{th:locrep}
		Let $U$ be a vertex-preserving unitary causal operator over space ${\cal H}$. Then,  in ${\cal H}'$ there exists $(K_u)$ such that for all $\ket{\psi}$:
		$$U\ket{\psi}=(\prod_{u\in V} \mu_u)\,(\prod_{u\in V} K_u)\,\ket{\psi}$$
		where $(K_u)$ is a collection of commuting unitary $n$-localized operators.
\end{Th} 
{\bf Proof.} Let us consider a causal operator $U$ over ${\cal H}$.
We define $K_u$ as $U'^\dagger \mu_u U'$, where $U'$ is a marked extension of $U$.
Using the causality of $U'$ and the dual causality property, we have that $K_u$ is a $n$-localized operator. Moreover, it is easy to see that for two distinct vertices $u$ and $v$:
\begin{align*}
 K_u K_v &=U'^\dagger \mu_u U'U'^\dagger \mu_v U'&\\
 &=U'^\dagger \mu_u  \mu_v U'&\\
 &=U'^\dagger \mu_v  \mu_u U'& \textrm{using commutativity of $\mu_\bullet$}\\
 &=U'^\dagger \mu_v U'U'^\dagger \mu_u U'&\\
 &=K_v K_u
\end{align*}
Hence, $(K_u)$ is a collection of localized commutating operators, thus the product $(\prod_u K_u)$ is well defined.\\
Now let us unfold the product $(\prod_u K_u)$ and apply it to $\ket{G}$, with $V(G)=\{u_1, \ldots, u_k\}$:
\begin{align*}
 K_{u_1} K_{u_2} \cdots K_{u_k} \ket{G} & =U'^\dagger \mu_{u_1} U' U'^\dagger \mu_{u_2} U'\cdots U'^\dagger \mu_{u_k} U'\ket{G}\\
 		&=U'^\dagger \mu_{u_1}  \mu_{u_2} \cdots  \mu_{u_k} U' \ket{G}\\
 		&=U'^\dagger \mu_{u_1}  \mu_{u_2} \cdots  \mu_{u_k}  U\ket{G}\\&\ \ \ \ \ \ \textrm{by construction of $U'$}\\
 		&= \mu_{u_1}  \mu_{u_2} \cdots  \mu_{u_k}  U\ket{G}
\end{align*}
Hence, applying $(\prod_u \mu_u)(\prod_u K_u)$ results in applying $(\prod_u \mu_u)^2 U$ which is just $U$.

The general implications of this Theorem are discussed in Secs \ref{sec:introduction} and \ref{sec:conclusion}. Mathematically speaking, the following two corollaries immediately follow.

\begin{Cor}[Inverse of unitary causal is causal]~\label{cor:inverse}
Let $U$ be a vertex-preserving unitary causal operator over space ${\cal H}$. Then,  $U^\dagger$ is also causal.
\end{Cor}
{\bf Proof outline.} The hypotheses imply localizability. The obtained circuit can then be reversed so as to implement $U^\dagger$. It follows that $U^\dagger$ is localizable, and hence causal. 

\begin{Cor}[Unitary 1-causal is causal]~\label{cor:1causality}
An operator $U$ over $\joliH$ is said to be 1-\emph{causal} if and only if it is vertex preserving and there exists $n\geq 0$ such that for any state $\rho$ over $\joliH$, and for any $v\in V$, we have
\begin{align}
(U\rho U^\dagger)^1_v=(U\rho^n_v U^\dagger)^1_v. \label{eq:1causality}
\end{align}
Let $U$ be a vertex-preserving unitary $1$-causal operator over space ${\cal H}$. Then, $U$ is causal.
\end{Cor}
{\bf Proof outline.} By inspection of the proofs above, it suffices to be $1$-causal in order to be localizable. But then localizability implies causality.

Notice that we could have followed the same reasoning for $0$-causality if the mark operator could have been made $0$-local, but avoiding port conflicts forces it to be $1$-local. Ultimately the demand to avoid port conflict is to ensure that the graphs remain bounded degree at all times. This in turn is just a choice: with unbounded degree graphs \cite{QuantumGraphity1,QuantumGraphity2} this question would not arise, and $0$-causality would imply causality. But then these graphs would be much harder to interpret geometrically, as dual to pseudo-manifolds \cite{ArrighiSURFACES} or spin networks.

\section{Summary and future work}\label{sec:conclusion}

We took as our state space the Hilbert space of quantum superpositions of finite, undirected, bounded-degree, labelled graphs. We adapted the notions of tensor product, partial trace, causality and locality to this context, mainly through the idea that they should coincide with their traditional counterparts whenever the graph is not in a superposition---and extending from there by linearity. We recovered a number of reassuring results, such as the duality between causality phrased in the Shr\"odinger picture (i.e. $(U\rho U^\dagger)^m_v=(U\rho^n_v U^\dagger)^m_v$) and that phrased in the Heisenberg picture (i.e. if $A$ is localized, so is $U^\dagger A U$), and a similar duality for localizabity.\\
Mainly, we showed that any vertex-preserving, global unitary causal operator $U$, decomposes into a finite-depth circuit of local unitary gates $\mu$ and $K$:
	$$U\ket{\psi}=(\prod  \mu_u)(\prod K_u)\ket{\psi}$$
thereby making the model constructive and parametrizable. Here $\mu$ is just a fixed, marking operator: only the $K_u$'s depend upon $U$. These commute: $[K_u,K_v]=0$. Hence the first product, which is the relevant one, can be spectrally decomposed. 

Here are two items for future work:
\begin{itemize}
\item The structure theorem was proven under the assumption that the global evolution is vertex-preserving. We wish to understand whether this condition can be relaxed. For instance, even in the model as it stands, we could attach to each vertex some `reservoir' of vertices, structured as an infinite binary tree. The quantum causal graph dynamics would then be able to `create' vertices by pulling them out of the reservoir, and to `destroy' them by pushing them back in. We plan to investigate this question in the near future.
\item Our framework is canonical, in the sense that it relies upon a distinguished discrete-time evolution. It cannot, therefore, be manifestly covariant in the sense of general relativity---but the perhaps covariance of some instances could be proven, e.g. in the style of \cite{arrighi2014discrete}. 
\end{itemize}

\section{Quantum gravity landscape}

\noindent {\em Digital Physics}, of which a prominent actor is \cite{tHooftCA}, seeks to recover modern theoretical physics concepts as emergent from a Reversible Cellular Automata. The above-presented Quantum Causal Graph Dynamics clearly arises as a two-fold extension of Reversible Cellular Automata --- an extension to dynamical graphs on the one hand, and to quantum theory on the other hand. In this sense it shares common origins with the digital physics program, but it also departs from it: both quantum dynamics and geometrodynamics \cite{WheelerGeo} are seen as fundamental features that one needs to put in. 

{\em Quantum Graphity} \cite{QuantumGraphity1,QuantumGraphity2} discretizes `pre-geometrical space' as a complete graph, with qubits on each edge telling whether its end vertices are neighbours or not. These qubits are then made to evolve according to a nearest-neighbours Hamiltonian. Quantum Graphity thus does not place space and time on a equal footing, as one is discrete and the other continuous. It follows that, strictly speaking, after any finite period of time, information has propagated everywhere: causality is approximate, in the sense of a Lieb-Robinson bound \cite{EisertSupersonic}. Quantum Causal Graph Dynamics can be seen as a theory of discrete-time quantum graphity.  

{\em Quantum Causal Histories} \cite{FotiniQCH}, on the other hand, does look at global unitary evolutions between two spacelike surfaces---but the fact that it decomposes into local unitary gates is directly assumed in this approach. The local unitary gates are also located at each vertex, and act upon the quantum information circulating upon the edges---but have no influence upon the graph itself. The causal set is given, and in no superposition.

{\em Emergence of almost-flat space.} In quantum graphity, the underlying graph is complete, which makes its geometrical interpretation very difficult. Some bounded-degree graphs, on the other hand, can be understood as pseudo-manifolds \cite{Lickorish,GrasselliGems,ArrighiSURFACES}.
Thus a more long-term aim is to retake this inspiring program of emergence of almost-flat space \cite{QuantumGraphity1,QuantumGraphity2} but in this discrete-space discrete-time bounded-degree-graphs formalism. Alternatively and interestingly, almost-flatness can also be seen as an emergent property of certain probability distribution over graphs \cite{TrugenbergerEmergent1,TrugenbergerEmergent2}, e.g. via clustering \cite{KrioukovEmergent}.

{\em Loop Quantum Gravity} \cite{RovelliLQG}, one of the main contenders for a theory of Quantum Gravity, also provides means of computing the transition amplitude of one space-like graph evolving into another. Yet its relationship with the above-presented Quantum Causal Graph Dymanics is unclear to us at this stage. Indeed, in Loop Quantum Gravity, the transition amplitude between the two spin networks, that make up the past and future boundaries of a spacetime region, is provided in a path-integral form by summing over the possible spin foams that could relate them. It follows that: the `time taken for the transition to happen' is also summed over; that the evolution is not guaranteed to be unitary; and that far-away vertices of the spin networks are allowed to signal to some extent. Actually, spin networks are not directly interpretable as discretized space-like surfaces in Loop Quantum Gravity: only coherent superpositions of them correspond to piecewise-linear manifolds in a one-to-one manner. These are some of the key differences that stand in the way of bridging this gap, which we believe would be a fruitful program.

{\em Causal Dynamical Triangulations} \cite{LollCDT} lets some of these difficulites vanish. It lifts the ambiguity of spin networks and directly works with glued equilateral tetrahedras. It works out the transition amplitude between two discretized space-like space surfaces that are separated by a given proper time. It does so in a path-integral form, by summing over the possible successive surfaces that could relate them --- but these  successive surfaces are themselves related by local 'moves'. Still it is unclear whether this induces a  unitary evolution operator over discretized space-like space surfaces. Moreover here again and far-away parts of the triangulation are allowed to signal to some extent.

Summarizing, quantum causal graph dynamics provide a mathematical framework for discrete-time quantum gravity models that exhibit both strict causality and strict unitary. To the best of our knowledge, to this day none of the main quantum gravity models gathers all of these features at once. Several of them are not so far. Gathering the reminding features would make the model fall within the scope of the structure theorem of this paper, and therefore decompose into local unitary scattering matrices $K$. This would pave the way towards quantum simulating Quantum Gravity.

\section*{Acknowledgements} This work has been funded by the ANR-12-BS02-007-01 TARMAC grant, the ANR-10-JCJC-0208 CausaQ grant, the John Templeton Foundation grant ID 15619, the STICAmSud project 16STIC05 FoQCoSS. The authors acknowledge enlightening discussions with Marios Christodoulou, Gilles Dowek and Simon Perdrix. 

\bibliography{biblio}
\bibliographystyle{plain}

\end{document}